\documentclass[epj]{svjour}
\usepackage{graphics}
\begin{document}
\title{SU(4) symmetry of the dynamical QCD string and
genesis of hadron spectra}

\author{L. Ya. Glozman  
}                     
\institute{Institute of Physics,  University of Graz, A--8010 Graz, Austria}
\date{Received: date / Revised version: date}
%
\abstract{
A large degeneracy of mesons of a given spin has recently been discovered 
upon reduction of the quasi-zero modes of the Dirac operator in a dynamical
lattice simulation. Here it is shown that a symmetry group
$SU(4) \supset SU(2)_L \times SU(2)_R \times U(1)_A \times {\cal C}_i$
is consistent with the observed degeneracy.
It is argued that this symmetry group is a symmetry of the dynamical
QCD string. Implications of this picture for a genesis of light hadron spectra
are discussed.
} 
\maketitle
\section{Introduction}
\label{intro}
A large degeneracy of
mesons of a given spin  has recently been discovered
in a dynamical lattice simulation
upon reduction of the lowest-lying eigenmodes of the manifestly
chirally-invariant overlap Dirac operator from the quark
propagators \cite{DGL1,DGL2} (for a
previous lattice study with the not chirally-invariant Wilson Dirac
operator see Refs. \cite{LS,GLS}).
Of course, after such a truncation the correlators do not correspond
to a local quantum field theory.\footnote{A nonlocality turnes out to be very
small, however, because a tiny amount of modes is removed, of order 10 from
more than one million.} Despite that fact the correlators turned out to be
very interesting: They have demonstrated a very clean exponential decay
for all $J=1$ channels and showed intriguing symmetry patterns.
A similar
degeneracy is seen in the observed highly excited mesons \cite{G1,G11}.
 
 The quasi-zero eigenmodes of the Dirac
operator are directly related to the chiral symmetry breaking quark condensate
via the Banks-Casher relation \cite{BC}. Consequently, if  hadrons
survive this artificial restoration ("unbreaking") of chiral symmetry
one expects that hadrons should fall into chiral multiplets. 

The complete set of all possible $\bar q q$ chiral multiplets of
the $J=1$ mesons is given in Table \ref{tab:t1} \cite{G1,G11}.
\begin{table}
\caption{The complete set of $\bar q q$ $J=1$ states 
classified according to the chiral basis. 
The symbol $\leftrightarrow$ indicates the states belonging to 
the same representation $r$ of the parity-chiral group
that must be degenerate in the $SU(2)_L \times SU(2)_R$ symmetric world.
Mesons belonging to the singlet representation (0,0) have no chiral partners.}
\label{tab:t1}
\begin{tabular}{ll}
\hline\noalign{\smallskip}

$r$  & mesons\\
\noalign{\smallskip}\hline\noalign{\smallskip}
$(0,0)$&$\omega(I=0,1^{--})~;~  f_1(I=0,1^{++})$\\
$(1/2,1/2)_a$&$\omega(I=0,1^{--}) \leftrightarrow b_1(I=1,1^{+-})$\\
$(1/2,1/2)_b$&$h_1(I=0,1^{+-}) \leftrightarrow \rho(I=1,1^{--}) $\\
$(0,1) \oplus (1,0)$&$a_1(I=1,1^{++} )\leftrightarrow \rho(I=1,1^{--})$\\
\noalign{\smallskip}\hline
\end{tabular}
\end{table}

Upon unbreaking of the chiral symmetry the states within each independent
chiral multiplet get degenerate. However, what is completely unexpected,  
not only a degeneracy within
chiral multiplets is seen, but actually a degeneracy of all eight $J=1$
mesons.  This degeneracy is obviously not accidental and tells us
something important about
the underlying dynamics.

A degeneracy  of four mesons from the $(1/2,1/2)_a$ and $(1/2,1/2)_b$
representations indicates a restoration of the $SU(2)_L \times SU(2)_R \times
U(1)_A$ symmetry \cite{G1,G11}. This symmetry does not connect, however,
these four mesons with other mesons from  Table \ref{tab:t1}. Consequently,
a degeneracy of all mesons from  Table \ref{tab:t1} implies a larger symmetry,
that includes $SU(2)_L \times SU(2)_R \times U(1)_A$ as a subgroup. Our primary
purpose is to establish this new symmetry.  Given this new symmetry we discuss
the physics implications for the highly degenerate system that is observed
and various ramifications, in particular a genesis of light hadron spectra.

\section{Parity-chiral $\bar q q$ multiplets.}
\label{sec:1}

In order to proceed we need to construct the 
parity-chiral $\bar q q$ multiplets of states of any spin.
The chirally symmetric $\bar q q$ states can be specified with the
following set of quantum numbers: ${r;IJ^{PC}}$, where $r$ is an index
of the parity-chiral group and all  other quantum numbers are
isospin ($I$), spin ($J$), spatial and charge parities ($P$ and $C$).
The $\bar q q$ states with $J \geq 1$ fill out the following possible
irreducible representations of the parity-chiral group $SU(2)_L \times SU(2)_R \times {\cal C}_i$,
where a group ${\cal C}_i$ consists of the space inversion and identity transformation
(a product with this group is required to construct states of definite parity):

{\bf (i)~~~ (0,0):}

\begin{equation}
|(0,0); \pm; J \rangle = \frac{1}{\sqrt 2} |\bar R R \pm \bar L L\rangle_J.
\label{00}
\end{equation} 

\noindent
Here the isospin $I=0$, $R$  denotes the right-handed fundamental $SU(2)_R$
vector, $R^T=(u_R,d_R)$, while $L$ describes the left-handed
 $SU(2)_L$ one,  $L^T=(u_L,d_L)$.
The index $J$ means that a definite spin $J$ and its projection $M$
are ascribed
to the given quark-antiquark system according to the 
relativistic spherical helicity formalism \cite{JW,LL}:

\begin{equation}
|\lambda_q\lambda_{\bar q}\rangle_J=D^{(J)}_{\lambda_q - \lambda_{\bar q}, M}(\vec n)\sqrt {\frac{2J+1}{4\pi}}
|\lambda_q\rangle  |-\lambda_{\bar q}\rangle,
\label{hel}
\end{equation}
where $D^{(J)}_{MM'}(\vec n)$ is the standard Wigner $D$--function describing rotation from the quantization axis to
the quark momentum direction $\vec n=\vec{p}/p$ and 
$\lambda_{q}$ ($\lambda_{\bar q}$) are the quark (antiquark) helicities;
 the quark chirality and helicity coincide, while for the antiquark
 they are just opposite. The parity of the quark-antiquark state is then given
 as

\begin{equation}
\hat P |(0,0); \pm; J \rangle = \pm (-1)^J  |(0,0); \pm; J \rangle.
\label{P00}
\end{equation}

{\bf (ii)~~~ $(1/2,1/2)_a$ and $(1/2,1/2)_b$:} 

\begin{equation}
|(1/2,1/2)_a; +;I=0; J \rangle = \frac{1}{\sqrt 2} |\bar R L + \bar L R\rangle_J,
\label{I0+1}
\end{equation} 

\begin{equation}
|(1/2,1/2)_a; -;I=1; J \rangle = \frac{1}{\sqrt 2} |\bar R \vec \tau L -
\bar L \vec \tau R\rangle_J,
\label{I0+2}
\end{equation} 

\noindent
and

\begin{equation}
|(1/2,1/2)_b; -;I=0; J \rangle = \frac{1}{\sqrt 2} |\bar R L - \bar L R\rangle_J,
\label{I0-1}
\end{equation}

\begin{equation}
|(1/2,1/2)_b; +;I=1; J \rangle = \frac{1}{\sqrt 2} |\bar R \vec \tau L +
\bar L \vec \tau R\rangle_J.
\label{I0-2}
\end{equation}

\noindent
In these expressions $\vec \tau$ are isospin Pauli matrices.
The parity of  every state in these representations is determined as

\begin{equation}
\hat P |(1/2,1/2); \pm; I; J \rangle = \pm (-1)^J |(1/2,1/2); 
\pm; I; J \rangle.
\label{P12}
\end{equation}

Note that a sum of the two distinct $(1/2,1/2)_a$ and $(1/2,1/2)_b$
 irreducible representations
of $SU(2)_L \times SU(2)_R$ forms an irreducible representation
of the $U(2)_L \times U(2)_R$ or $SU(2)_L \times SU(2)_R \times U(1)_A$ groups.

{\bf (iii)~~~ (0,1)$\oplus$(1,0):} 

\begin{equation}
|(0,1)+(1,0); \pm; J \rangle = \frac{1}{\sqrt 2} |\bar R \vec \tau R 
\pm \bar L  \vec \tau L \rangle_J,
\label{10}
\end{equation} 

\noindent
the isospin $I=1$ and   parities

\begin{equation}
\hat P |(0,1)+(1,0); \pm; J \rangle = \pm (-1)^J 
 |(0,1)+(1,0); \pm; J \rangle.
\label{P10}
\end{equation} 

For the $J=0$ states the representations $(0,0)$ and $(0,1)+(1,0)$
are impossible, because the total spin projection onto the momentum
direction of the quark for these representations is $\pm 1$.

All the basis vectors (1),(4-7) and (9) are the relativistic spherical
helicity states in the quark-antiquark system  
that represent natural relativistic 
basis for the
bound quark-antiquark states with definite chirality.
They should not be
confused with the plane waves. They are correct relativistic
basis vectors that carry
complete information about chiral degrees of freedom 
in a meson with restored chiral symmetry and thus should be sufficient
to reconstruct the observed higher symmetry.

One can also construct various local \cite{CJ} and nonlocal \cite{G1,G11}
composite $\bar q q$ operators that
have the required chiral symmetry properties.

\section{The $SU(4)  \supset SU(2)_L \times SU(2)_R
\times U(1)_A \times {\cal C}_i$ symmetry.}
\label{sec:2}

Our task is to find a minimal symmetry group that combines
all four irreducible representations $(0,0)$, $(1/2,1/2)_a$, $(1/2,1/2)_b$
 and $(0,1)+(1,0)$ of the parity-chiral group into one 
 representation of a larger group. Transformations of this group should
 connect all basis vectors (1), (4-7) and (9) to each other.
 
 Transformations that link (4) with (5) and (6) with (7) are the
 $SU(2)_L \times SU(2)_R$ transformations, i.e.,  independent
 rotations of both right-handed and left-handed fundamental vectors
 $R$ and $L$ in the isospin space. In order to connect (4-5) with (6-7)
 we need in addition the $U(1)_A$ transformation, that links the
 $(1/2,1/2)_a$ and $(1/2,1/2)_b$ states of the same isospin but opposite
 parity. The $SU(2)_L \times SU(2)_R \times U(1)_A$ transformations do not connect,
 however, (4-7) with (1) or (9), because both the basis vectors (1) and (9)
 are selfdual with respect to $U(1)_A$. Consequently, in order to find
 a symmetry group that connects all basis vectors (1), (4-7), (9), we
 need to find  transformations that link the states (4-7) with (1) and (9).
 
 Such  transformations can be most transparently seen when we use explicit 
 notations for the basis
 vectors. Consider, as an example, the $Q=-1$ charge states of (7) and (9) of
 equal parity:
 
 \begin{equation}
\frac{1}{\sqrt 2}|{\bar u}_R d_L + {\bar u}_L d_R\rangle_J ~~~and~~~
\frac{1}{\sqrt 2}|{\bar u}_R d_R + {\bar u}_L d_L\rangle_J.
\end{equation}

\noindent
A symmetry transformation that connects both these states is
$(d_L \leftrightarrow d_R) \otimes (u_L \leftrightarrow u_L) 
\otimes (u_R \leftrightarrow u_R)$. This cannot be a parity transformation,
because the space inversion transforms the left-handed quarks into the
 right-handed
quarks and
vice versa for both flavors simultaneously. Such a transformation can be
obtained if we perform two independent $SU(2)_U$ and $SU(2)_D$ rotations
of two independent fundamental vectors $U$ and $D$,
 where $U^T=(u_L,u_R)$ and $D^T=(d_L,d_R)$. Similarly, the $Q=+1$ states
 of (7)
 and (9) of the same parity
 
 \begin{equation}
\frac{1}{\sqrt 2}|{\bar d}_R u_L + {\bar d}_L u_R\rangle_J ~~~and~~~
\frac{1}{\sqrt 2}|{\bar d}_R u_R + {\bar d}_L u_L\rangle_J,
\end{equation}

\noindent
transform into each other through
$(u_L \leftrightarrow u_R) \otimes (d_L \leftrightarrow d_L) 
\otimes (d_R \leftrightarrow d_R)$, which again can  be accomplished
via two independent $SU(2)_U$ and $SU(2)_D$ rotations.

One can check that the same is true for the $Q=0$ states of (7) and (9) as well
as for the $Q=0$ states (1) and  (4).

Now we are in a position to find a minimal symmetry group
that connects all vectors (1),(4-7) and (9). This group must contain
as  subgroups the $SU(2)_L$ and $SU(2)_R$ isospin rotations of quarks
of fixed chirality, the $SU(2)_U$ and $SU(2)_D$ chirality rotations
of quarks with  fixed flavor, the  $U(1)_A$, as well as a parity
transformation $(u_L \leftrightarrow u_R) \otimes (d_L \leftrightarrow d_R)$.
This symmetry transforms
 the fundamental four-component vector $N$, $N^T =(u_L,u_R,d_L,d_R)$
and represents the $SU(4)$ group. Vectors (1),(4-7) and (9) form a basis set
for a  dim=16 reducible representation $ \bar {4} \times {4}
= {15} + {1}$   of the group $SU(4)$
in the reduction chain $SU(4)  \supset SU(2)_L \times SU(2)_R
 \times {\cal C}_i$. 

An important issue is that this new $SU(4)$ symmetry
is relevant only to $J \geq 1$ states. For the $J=0$ states only
the basis vectors (4-7) are possible and the total symmetry group
that combines all possible states of the $J=0$ mesons is   
$SU(2)_L \times SU(2)_R
\times U(1)_A $. 

We stress that this symmetry is not a symmetry of the QCD Lagrangian.
It should be considered as an emergent symmetry that appears from the QCD
dynamics upon removal of the quasi-zero modes of the Dirac operator.

The  ultra-relativistic $SU(4)  \supset SU(2)_L \times SU(2)_R
\times U(1)_A \times {\cal C}_i$ symmetry should not 
be confused with the nonrelativistic
Wigner spin-isospin \cite{W,GR} (heavy-quark \cite{IW1,IW2}) $SU(4)_{SI} \supset SU(2)_S \times SU(2)_I$ 
symmetry.  It should
also not be confused with the $SU(4)$ Pauli-G\"{u}rsey symmetry \cite{P,G} that connects
the mesonic and baryonic (diquark) states within  $N_c=2$ QCD.

Finally, we note that a generalization of this symmetry to  $N_f$
light flavors is straightforward and the relevant symmetry group in
this case is $SU(2N_f)$.

\section{Genesis of light meson spectra.}
\label{sec:3}

 Within the potential
constituent quark model \cite{RGG,GI}, that was a basis for intuition and
insights for many years, a gross symmetry of the light hadron spectra
is $SU(4)_{SI}\times O(3)$, which is a symmetry of the levels of the
confining interquark potential. This symmetry gets broken by the phenomenologically
introduced spin-spin, tensor and spin-orbit interactions that are fitted
to the experimental levels. As a consequence the $SU(4)_{SI}$ symmetry is
lifted. Such a physical picture has a solid basis in the heavy quark mesons
but cannot be substantiated in the light quark sector where chiral and
$U(1)_A$ symmetries and their breakings are crucially important.

The results presented above suggest that the primary energy
level has a symmetry $SU(4)  \supset SU(2)_L \times SU(2)_R
\times U(1)_A \times {\cal C}_i$, not to be confused with the nonrelativistic
$SU(4)_{SI}$ symmetry of the constituent quark model. E.g., the former 
symmetry combines into one multiplet of
dim=16 all mesons from  Table \ref{tab:t1}, while a dim=16 multiplet
 of $SU(4)_{SI}$ consists of the $\pi, \eta_2, \rho, \omega$ mesons.
The primary energy levels observed in \cite{DGL1,DGL2} contain all degenerate states of
both parity from  Table \ref{tab:t1}, while the positive and the negative
parity levels of the confining potential of the
constituent quark model represent different $SU(4)_{SI}\times O(3)$ multiplets
that are strongly splitted (with the harmonic confinement
this splitting is $\hbar\omega$).

A genesis of the light quark $\bar q q$ mesons could be then viewed as
follows:
A confining interaction gives rise to the highly degenerate primary
levels with the symmetry $SU(4)  \supset SU(2)_L \times SU(2)_R
\times U(1)_A \times {\cal C}_i$ and a dynamics related to the quasi-zero modes
of the Dirac operator supplies a breaking of both chiral and $U(1)_A$
symmetries as well as a splitting of the primary confining levels.
While both  $SU(2)_L \times SU(2)_R$ and $U(1)_A$ breakings are most
probably related to the instanton-induced dynamics \cite{H,S,DP}, different
effective microscopic mechanisms could be at work for splitting of the primary energy levels.

\section{The dynamical QCD string.}
\label{sec:4}

In Ref. \cite{DGL1} it was speculated that the highly degenerate energy levels
observed after subtraction of the lowest Dirac eigenmodes  are quantum levels 
of the dynamical QCD string. Below we precisely
formulate  arguments that lead to such a conclusion.

Consider a motion of an electrically charged fermion in a static electric field
or a relative motion of two charged fermions. In such systems there exist
 magnetic interactions which manifest themself through the spin-spin,
spin-orbit and tensor interactions. In our case we do have a relative motion
of two color-charged fermions, however the spin-spin, spin-orbit and tensor
interactions are absent. This can be proved as follows.

All relativistic chiral states from  Table \ref{tab:t1} can be decomposed
via the unitary transformation into a sum of vectors of the
$\{I, ^{2S+1}L_J\}$ basis \cite{GN1,GN2}:

\begin{eqnarray}
\displaystyle |(0,1)+(1,0);1 ~ 1^{--}\rangle&=&\sqrt{\frac{2}{3}}\,|1;{}^3S_1\rangle+\sqrt{\frac{1}{3}}\,|1;{}^3D_1\rangle,\nonumber\\
\displaystyle |(1/2,1/2)_b;1 ~
1^{--}\rangle&=&\sqrt{\frac{1}{3}}\,|1;{}^3S_1\rangle-\sqrt{\frac{2}{3}}\,|1;{}^3D_1\rangle,\nonumber\\
\displaystyle |(0,0);0 ~ 1^{--}\rangle&=&\sqrt{\frac{2}{3}}\,|0;{}^3S_1\rangle+\sqrt{\frac{1}{3}}\,|0;{}^3D_1\rangle,\nonumber\\
\displaystyle |(1/2,1/2)_a;0 ~
1^{--}\rangle&=&\sqrt{\frac{1}{3}}\,|0;{}^3S_1\rangle-\sqrt{\frac{2}{3}}\,|0;{}^3D_1\rangle,\nonumber\\
\displaystyle |(0,1)+(1,0);1 ~ 1^{++}\rangle&=&|1;{}^3P_1\rangle,\nonumber\\
\displaystyle |(0,0);0 ~ 1^{++}\rangle&=&|0;{}^3P_1\rangle,\nonumber\\
\displaystyle |(1/2,1/2)_a;1 ~ 1^{+-}\rangle&=&|1;{}^1P_1\rangle,\nonumber\\
\displaystyle |(1/2,1/2)_b;0 ~ 1^{+-}\rangle&=&|0;{}^1P_1\rangle.\nonumber\\
\end{eqnarray}

We can invert this unitary transformation and obtain a chiral decomposition
of vectors

\begin{equation}
|0;{}^3S_1\rangle, |1;{}^3S_1\rangle,|0;{}^3D_1\rangle,|1;{}^3D_1\rangle, 
\label{LS1}
\end{equation}
\begin{equation}
 |0;{}^1P_1\rangle,|1;{}^1P_1\rangle,|0;{}^3P_1\rangle,|1;{}^3P_1\rangle.
\label{LS2}
\end{equation}

Given that all eight states from  Table \ref{tab:t1} are degenerate, we immediately obtain a degeneracy of all
eight states (\ref{LS1}-\ref{LS2}). This degeneracy implies absence of the spin-spin, spin-orbit and
tensor interactions in the system. Indeed, a nonzero spin-orbit force would split the
$^3S_1$ and $^3P_1$; the $^1P_1$ and $^3P_1$; etc. terms, a nonzero spin-spin force would split
the $^1P_1$ and $^3P_1$; the $^3S_1$ and $^1P_1$; etc levels, and a tensor force would split the
$^3S_1$ and $^3D_1$ terms. 

We conclude that there are no magnetic interactions in the system.
The energy of the system is entirely due to interactions of the color charges via the
color-electric field and due to a relativistic motion of the system. 
We interpret (or, better, define) such
a system as a dynamical QCD string (for a simple model see \cite{G2}). 

One might raise a question why such a system without the color-magnetic field
is interpreted by us as a dynamical QCD string? In QED in 3+1 dimensions
a motion of an electric charge in a vacuum induces a magnetic field that lives
in a plane that is perpendicular to the charge motion. With one spatial
dimension the system becomes, however, pure electric. The absence of the
color-magnetic field in our case is natural if the system became effectively
one-dimensional. A one-dimensional dynamical color-electric system with chiral
quarks at the ends, that is embedded into 3+1 dimensional space, is natural to
call a string. We do not know, however, why such a dimesional reduction should
happen in QCD with charges that move with the speed of light.

\section{Summary}
\label{sec:5}

We have suggested a new symmetry that is associated with
the degeneracy of the energy levels of mesons of a given spin $J \geq 1$
after subtraction of the quasi-zero eigenmodes of the Dirac
operator. It is $SU(4)  \supset SU(2)_L \times SU(2)_R
\times U(1)_A \times {\cal C}_i$. It is not a symmetry of the QCD Lagrangian,
but is an emergent symmetry that appears from the QCD dynamics upon reduction
of the low-lying Dirac modes. We consider this symmetry
as a symmetry of the confining
interaction in QCD. Actually the symmetry group could be even higher
if the energy levels of mesons with different spins will turn out to be
degenerate. The latter issue is a subject of the current lattice simulations.

We interpret these 
highly degenerate energy levels as levels of the dynamical QCD
string. We actually define such a system as the dynamical QCD string, because
there is no magnetic interaction in the system.  
The energy of the system comes
only from the color-electric interaction and from the relativistic motion
of the system. This picture should be contrasted with the well understood
relative motion of two fermions within the local $U(1)$-gauge theory where magnetic
interaction is necessarily present.

A genesis of the light meson spectra looks quite different as compared
to the constituent quark model. The 
$\bar q q$ spectra could be viewed as a result of the splitting of the primary
energy levels of the dynamical QCD string by means of dynamics associated
with the quasi-zero modes of the Dirac operator.

\medskip
 We thank C.B. Lang for discussions and careful reading of the
 manuscript. Partial support from the
Austrian Science Fund (FWF) through  grant
P26627-N27 is acknowledged.

\end{document}